# Molecular Beam Dumps as Tracers of Hadronic Cosmic Ray Sources: the Case of SNR IC443*

*based on a talk presented at the 2002 Moriond Gamma-Ray Universe Conference*

Yousaf M. Butt[1], Diego F. Torres[2], Jorge A. Combi[3], Thomas M. Dame[1], Gustavo E. Romero[3#]

[1]Harvard-Smithsonian Center for Astrophysics, Cambridge, MA, USA

[2]Department of Physics, Princeton University, Princeton, NJ, USA & Lawrence Livermore National Lab., CA, USA

[3]Instituto Argentino de Radioastronomía, C.C. 5, 1894, Villa Elisa, Buenos Aires, ARGENTINA

[#] presently at CEA/CNRS/Saclay, & Université de Paris, France

**The gamma-ray & neutrino visibility of cosmic ray (CR) accelerators will be dramatically increased by the presence of molecular material abutting such sites due to the increased probability of pion production – and, in the case of neutral pions, subsequent gamma-decay. This was already recognized by Montmerle, Pinkau, and Black & Fazio, and others in the 1970's. In an effort to examine the long-standing — but unproven — conjecture that galactic supernova remnants (SNRs) are indeed the sites of nucleonic CR acceleration to ≲300TeV/n we have carried out a 3-way coincidence search between the SNRs in Green's (2001) catalog, the GeV range unidentified sources from the third EGRET catalog (Hartman et al., 1999), and molecular clouds, at low Galactic latitudes ($|b|$≲5). In order to be *quantitative* regarding the distribution and amount of the ambient molecular masses we have extracted the CO($J_{rot}$=1→0) mm wavelength data from the compilation of Dame, Hartmann & Thaddeus (2001), at the best estimates of the various SNR distances. We outline the correlative study and examine the interesting case of SNR IC443, a likely accelerator of CR nuclei.**

Subject headings: acceleration of particles --- cosmic rays --- ISM: clouds --- supernova remnants --- IC443 --masers -- shock interaction

# INTRODUCTION

In 1953 Shklovskii speculated that *"it is possible that ionized interstellar atoms are accelerated in the moving magnetic fields connected with an expanding* [SNR] *nebula"* (Shklovskii, 1953). Despite the fact that there appears to be continuing broad consensus favoring this scenario, direct evidence confirming it has so far eluded observers. The situation has been reviewed by Luke Drury and collaborators, (2001); but also see the review by Rainer Plaga (2001) for alternative viewpoints. The best way to identify discrete energetic nucleonic CR accelerators, whatever their nature, is to look for the associated high-energy $\gamma$-rays − and, in the future, neutrinos (Halzen & Hooper, 2002) − produced at the source sites. However, since CR electrons and nuclei can both generate gamma-rays at the acceleration sites it has not yet been possible to unambiguously associate the detected gamma-radiation with sources of CR *nuclei,* in particular. (Neutrinos, too, can be generated non-hadronically [Athar, 2002].)

## MOLECULAR CLOUDS, CR ACCELERATORS & GeV SOURCES

Massive and dense molecular clouds lying adjacent to CR accelerators will dramatically increase the target density for the freshly accelerated nuclei, and thus increase the neutral (& charged) pion production: eg.

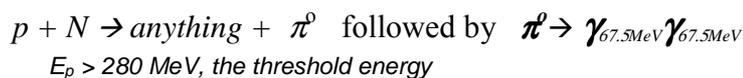

$$p + N \to anything + \pi^0 \quad \text{followed by} \quad \pi^0 \to \gamma_{67.5MeV}\gamma_{67.5MeV}$$
$E_p > 280$ MeV, the threshold energy

where N represents any nucleonic target (mostly protons). The outgoing $\pi^0$ will then — 98.8% of the time — decay isotropically to two 67.5 MeV ($=\frac{1}{2} m_{\pi^0}$) gamma-rays *in the pion rest frame,* $\pi^0 \to \gamma_{67.5MeV}\gamma_{67.5MeV}$. In the observer's frame these two 67.5 MeV gamma rays will be appropriately up *and* down shifted, depending on the pion velocity, and the gamma-ray direction (eg. Stecker, 1971). The overall observed gamma-ray spectrum from neutral pion decay will be peaked at 67.5 MeV and extend to lower and higher energies: the higher the energy of the parent protons (and thus the secondary pions), the broader the spectrum, as illustrated in Fig. 1 (taken from the superb article by Murphy, Dermer & Ramaty, 1987). It should be noted that the gamma-ray spectrum from a population of protons of maximum energy $E_{p\text{-}max}$, will be cut-off at about the same energy, $E_{\gamma\text{-}max} \sim E_{p\text{-}max}$ (eg. Naito & Takahara, 1994).

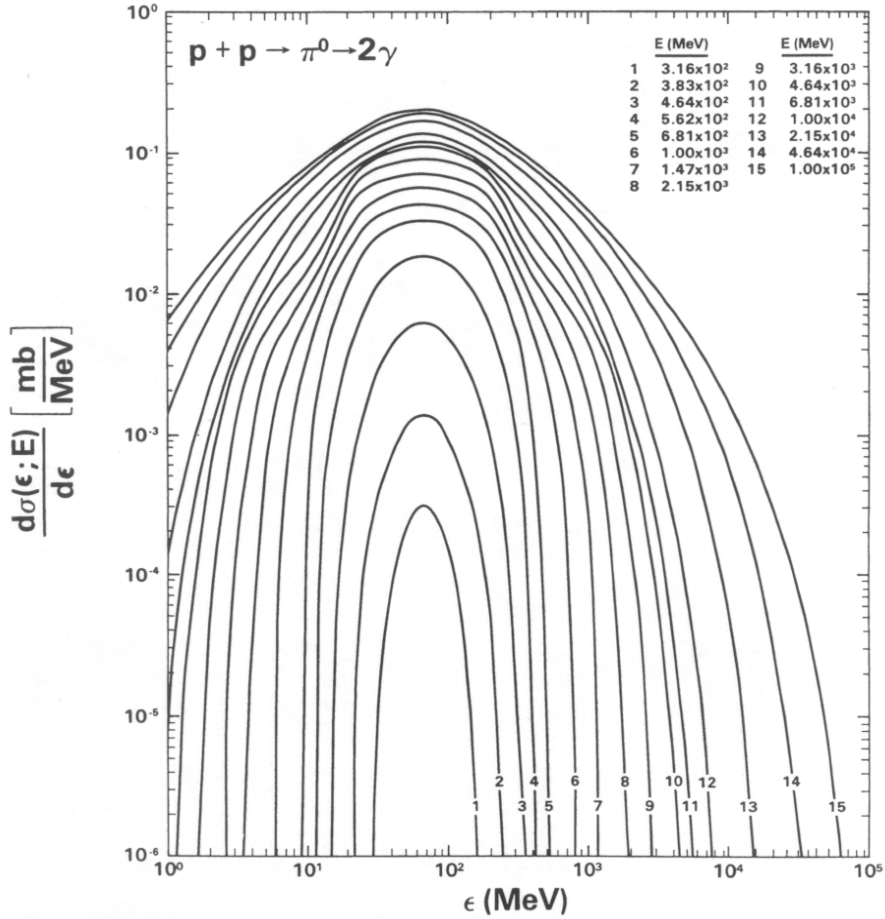

Fig 1: Energy spectra of gamma-rays resulting from the decay of $\pi^0$ mesons produced in collisions of isotropic, monoenergetic protons with protons at rest, at a variety of kinetic energies. *From Murphy, Dermer & Ramaty, 1987*. Note: in realistic astrophysical scenarios one also expects a contribution from the relativistic bremsstrahlung from the secondary leptons also produced in hadronic interactions (Schlickeiser 1982). eg. p + p → X + $\pi^\pm$, followed by $\pi^\pm$ → $\mu^\pm$ + $\nu$ → $e^\pm$ $\nu$ $\bar{\nu}$.

The role of molecular target material in enhancing the gamma-ray visibility of GCR sources has, of course, already been noted by many authors before us (eg. Black & Fazio, 1972; Montmerle, 1979; Dorfi, 1991, 2000; Aharonian, Drury & Volk, 1994).

We have carried out a search for 3-way coincidences between the locations of SNRs (Green 2001), 3EG unidentified sources (Hartman et al., 1999) at low galactic latitudes ($|b| \lesssim 5$), and the ambient molecular clouds as observed in the CO (1→0) mm wavelength rotational transition (Dame, Hartmann & Thaddeus, 2001), in an effort to test *quantitatively* the conjecture that SNRs accelerate nuclei to GCR energies. An example of this approach may be found in Butt et al., 2001.

First we identify all spatial coincidences between the SNRs in Green's (2001) catalog and unidentified EGRET sources (Hartman et al., 1999) at $|b| \lesssim 5$ (Table 1). Next, we examine on a case by case basis the molecular clouds (Dame, Hartmann & Thaddeus, 2001) at the distance of the SNRs, as well as other potential sources of the GeV radiation, such as nearby young, energetic pulsars, OB associations etc.

Table 1: Positional coincidences between SNRs quoted in the Green's Catalog (2001), and unidentified 3EG EGRET sources (Hartman et al., 1999) at $|b| \lesssim 5$. Class of the 3EG source is defined as: em → possibly extended; C→ confused. A "y" mark in the "P?" column implies that the 3EG source is also coincident with a radio pulsar in the Princeton or ATNF archives. $F_\gamma$ refers to the flux of the 3EG source above 100 MeV in units of $10^{-8}$ ph cm$^{-2}$s$^{-1}$. $\Gamma$ refers to the photon spectral index $F(E) \sim E^{-\Gamma}$. I is the variability index (Torres et al. 2001). $\tau$ is the Tompkins variability index (Tompkins 1999). $\Delta\theta$ is the positional offsets between the centers of the SNRs and the corresponding 3EG source, and the "Size" is the size of the remnant in arcmin. T is the type of SNR: S: shell-type, F: filled-center. C: composite.

| γ-source | $F_\gamma$ | $\Gamma$ | Class | I | $\tau$ | P? | SNR | Other name | $\Delta\theta$ | Size | T |
|---|---|---|---|---|---|---|---|---|---|---|---|
| 0542+2610 | 14.7±3.2 | 2.67±0.22 | em C | 3.16 | $0.70^{1.40}_{0.34}$ | | G180.0−1.7 | | 2.04 | 180 | S |
| 0617+2238[1,2] | 51.4±3.5 | 2.01±0.06 | C | 1.68 | $0.26^{0.38}_{0.15}$ | | G189.1+3.0 | IC443 | 0.11 | 45 | S |
| 0631+0642[1,3] | 14.3±3.4 | 2.06±0.15 | C | 1.52 | $75.8^{\infty}_{7.89}$ | | G205.5+0.5 | Monoceros | 1.97 | 220 | S |
| 0634+0521 | 15.0±3.5 | 2.03±0.26 | em C | 1.02 | $72.0^{\infty}_{0.15}$ | | G205.5+0.5 | Monoceros | 2.03 | 220 | S |
| 1013−5915 | 33.4±6.0 | 2.32±0.13 | em C | 1.63 | $0.22^{0.46}_{0.00}$ | y | G284.3−1.8 | MSH 10-53 | 0.65 | 24 | S |
| 1102−6103 | 32.5±6.2 | 2.47±0.21 | C | 1.86 | $0.00^{0.90}_{0.00}$ | | G290.1−0.8 | MSH 11-61A | 0.12 | 19 | S |
| | | | | | | | G289.7−0.3 | | 0.75 | 18 | S |
| 1410−6147[4] | 64.2±8.8 | 2.12±0.14 | C | 1.22 | $0.33^{0.55}_{0.16}$ | y | G312.4−0.4 | | 0.23 | 38 | S |
| 1639−4702 | 53.2±8.7 | 2.50±0.18 | em C | 1.95 | $0.00^{0.38}_{0.00}$ | y | G337.8−0.1 | Kes 41 | 0.07 | 9 | S |
| | | | | | | | G338.1+0.4 | | 0.65 | 15 | S |
| | | | | | | | G338.3+0.0 | | 0.57 | 8 | S |
| 1714−3857 | 43.6±6.5 | 2.30±0.20 | em C | 2.17 | $0.15^{0.38}_{0.00}$ | y | G348.5+0.0 | | 0.47 | 10 | S |
| | | | | | | | G348.5+0.1 | CTB 37A | 0.50 | 15 | S |
| | | | | | | | G347.3−0.5 | RX J1713-3946 | 0.85 | 65 | S  *TeV emitter* |
| 1734−3232[5] | 40.3±6.7 | – | C | 2.90 | $0.00^{0.24}_{0.00}$ | | G355.6+0.0 | | 0.16 | 8 | S |
| 1744−3011 | 63.9±7.1 | 2.17±0.08 | C | 1.80 | $0.38^{0.62}_{0.20}$ | | G359.0−0.9 | | 0.41 | 23 | S |
| | | | | | | | G359.1−0.5 | | 0.25 | 24 | S |
| 1746−2851[6] | 119.9±7.4 | 1.70±0.07 | em C | 2.00 | $0.50^{0.69}_{0.36}$ | | G0.0+0.0 | Galactic Center Region | 0.12 | 3.5 | S |
| | | | | | | | G0.3+0.0 | | 0.19 | 16 | S |
| 1800−2338[1,7] | 61.3±6.7 | 2.10±0.10 | C | 1.60 | $0.03^{0.32}_{0.00}$ | y | G6.4−0.1 | W28 | 0.17 | 42 | C |
| 1824−1514 | 35.2±6.5 | 2.19±0.18 | C | 3.00 | $0.00^{0.51}_{0.00}$ | y | G16.8−1.1 | | 0.43 | 30 | – |
| 1837−0423 | <19.1 | 2.71±0.44 | C | 5.41 | $12.0^{\infty}_{2.17}$ | y | G27.8+0.6 | | 0.58 | 50 | F |
| 1856+0114[8] | 67.5±8.6 | 1.93±0.10 | em C | 2.92 | $0.80^{1.31}_{0.50}$ | y | G34.7−0.4 | W44 | 0.17 | 35 | S |
| 1903+0550[4] | 62.1±8.9 | 2.38±0.17 | em C | 2.28 | $0.35^{0.60}_{0.18}$ | y | G39.2−0.3 | 3C396, HC24 | 0.41 | 8 | S |
| 2016+3657 | 34.7±5.7 | 2.09±0.11 | C | 2.06 | $0.37^{0.75}_{0.08}$ | | G74.9+1.2 | CTB 87 | 0.26 | 8 | F |
| 2020+4017[1,9] | 123.7±6.7 | 2.08±0.04 | C | 1.12 | $0.07^{0.18}_{0.00}$ | | G78.2+2.1 | W66, γ-Cygni | 0.15 | 60 | S |

[1] Association proposed by Sturner & Dermer (1995), and Esposito et al. (1996). [2] GeV J0617+2237 [3] GeV J0633+0645. [4] Association proposed by Sturner & Dermer (1995). [5] GeV J1732−3130. [6] GeV J1746−2854. [7] GeV J1800−2328. [8] GeV J1856−0115. [9] GeV J2020+4023. GeV sources compiled in the GeV ASCA Catalog (Roberts et al. 2001).

*From Torres et al., to appear in Phys. Rep., 2002*

We certainly are **_not_** proposing that each of these spatial coincidences necessarily implies a physical connection between the SNR and the EGRET source. For instance, in many cases there are pulsars which are spatially coincident with the 3EG sources, which complicate the assessment of the relevant SNR as the unique source of the observed GeV radiation (eg. Thompson, 2001). Indeed, it is likely that many of the unidentified 3EG sources at $|b|\lesssim5$ are truly composites given the EGRET spatial resolution.

The total pion produced gamma-ray luminosity of a CR source embedded in (or adjacent to) a dense medium can be divided between that from the hadronic interactions intrinsic to the source, and that due to the enhanced probability of hadronic interactions in the high target density medium of the ambient clouds: $F_{tot}(E>100\text{MeV}) = F_{snr} + F_{cloud}$

In the simple model of Drury et al (1994), we may evaluate the first term as:

$$F_{snr}(E>100\text{MeV}) \sim 4.4 \times 10^{-7} \, \theta \, E_{51} \, D^{-2}_{kpc} \, n_o \quad \text{photons cm}^{-2} \text{ sec}^{-1} \quad [1]$$

where $\theta$ is the fraction of the total supernova energy converted to cosmic ray energy; $E_{51}$ is the supernova explosion energy in units of $10^{51}$ erg; and $D_{kpc}$ is the distance in kpc. The second term in Eq. 1 represents the contribution to the gamma-ray flux from the SNR amplified CR bombardment of the adjacent clouds and can be approximated by (Aharonian & Atoyan, 1996):

$$F_{cloud\,A}(E>100\text{MeV}) = 2.2 \times 10^{-7} \, M_5 \, D^{-2}_{kpc} \, k_s \quad \text{photons cm}^{-2} \text{ sec}^{-1} \quad [2]$$

where $M_5$ is the mass in units of $10^5 \, M_o$ and $k_s$ is the cosmic ray enhancement factor, ie. the ratio of the CR energy density in the vicinity of the SNR to that in the solar system.

## THE INTERACTING SNR IC443

The case-by-case analyses for the SNRs in Table 1 will be presented in a forthcoming review article in *Physics Reports* by Torres et al., (2002). Here, by way of illustration, we briefly outline the interesting story of IC443. This was already proposed as a possible CR accelerator by Sturner & Dermer (1995); Sturner, Dermer & Mattox (1996); Esposito et al., (1996); Keohane et al., (1997), Gaisser, Protheroe & Stanev (1998) and Hnatyk & Petruk (1998), based on the coincidence of the bright GeV source 2EG J0618+2234 with the remnant. However, Olbert et al. (2001) then found a pulsar in their *CHANDRA* data in a region of hard x-ray emission close to the 95% confidence location contours of the 2EG source. But the position of the $\gamma$-ray source itself was also slightly modified in going from the 2EG (Thompson et al., 1995) to the 3EG catalog (Hartman et al., 1999), such that the pulsar and its wind nebula is no longer even close to the 95% confidence location contours of the source 3EG J0617+2238 ($\equiv$2EG J0618+2234). When we extracted the CO data for the remnant, we observed that there is a large amount of molecular mass ($\sim10^4 \, M_o$) consistent with the distance to this remnant (corresponding to a "local standard of rest" velocity, $v_{lsr}$=-40 to +20 km/sec) − see

Figure 2. This then resuscitates the original proposal that this SNR is likely a nucleonic CR accelerator and that the GeV emission is mostly due to pion decay. This interpretation is supported by the high energy spectrum of this source, which is suggestive of the characteristic pion "hump" feature (Fig. 3). The presence of shocked maser emission (Claussen et al., 1997) from $(l,b) \sim (189, 2.91)$ – the "⊗" shape in Figure 2, coincident with the γ-ray source – lends further support to this hypothesis.

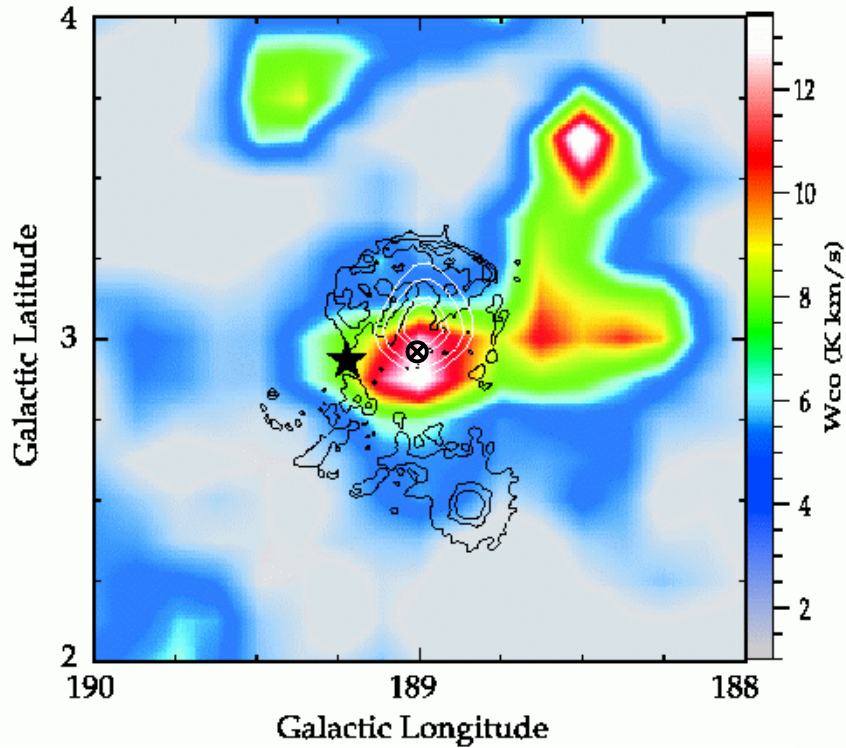

Fig. 2: The SNR IC443 is shown in the dark optical contours from Lasker et al. (1990). The distribution of the CO molecular mass in the velocity interval v=−40 to +20 km/sec is displayed in colour. We estimate the equivalent $H_2$ mass to be about ~7900 $M_o$, at the region towards the center of the remnant. The density of that cloud is about ~840 nucleons $cm^{-3}$. The EGRET source 3EG J0617+2238 is shown in white contours. The pulsar, CXOU J061705.3+222127 discovered by Olbert et al., 2001 is shown by the black star (★) shape. The EGRET source is *not* consistent with the location of the pulsar, nor with that of the hard X-ray emission of the plerion associated with the x-ray pulsar (Bocchino & Bykov, 2001). It is consistent with the location of the massive molecular cloud and the shocked maser emission reported by Claussen et al., (1997): the '⊗' shape in the adjacent figure.

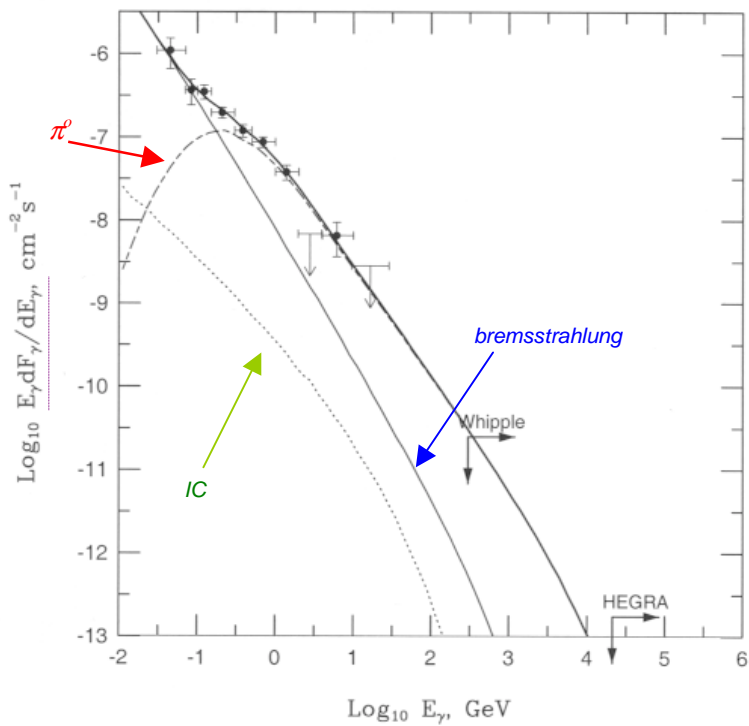

Fig. 3: A possible fit to the EGRET spectrum of IC443 *from Gaisser, Protheroe & Stanev (1998)*. The short-long dashed curve is the neutral pion contribution to the gamma-ray spectrum. The thin solid line is the electronic bremsstrahlung component, and the dotted curve is the contribution from electronic Inverse Compton off of the microwave background photon field. Several models were explored by Gaisser, Protheroe & Stanev (1998) to explain the GeV emission from IC443; the fit shown in the adjacent figure is the one for which the input parameters most closely match the density of the cloud we have derived from our CO extraction, and could, thus, be considered the most realistic model. (Note the units of the y-axis.)

Bykov et al., (2000) had suggested that the GeV emission seen towards IC 443 is due mostly to relativistic bremsstrahlung of $e^-$'s rather than to pion deacy via hadronic interactions of nuclei. Unfortunately, these authors had associated the synchrotron radio emission seen mostly towards the rim of the remnant with the centrally located EGRET source in their multiwavelength fits. It is now known that these two (synchrotron radio & $\gamma$-ray) emission regions are spatially well-distinct (Fig. 1) and thus cannot be combined arbitrarily in multiwavelength fits. This compromises the premise of Bykov et al.(2000) that $e^-$'s are favored over protons as the source of the $\gamma$-ray emission. (Certainly, there will be a weak component of non-thermal bremsstrahlung from secondary leptons also generated in hadronic interactions: eg. p + p → X + $\pi^\pm$, followed by $\pi^\pm \rightarrow \mu^\pm + \nu \rightarrow e^\pm \nu \bar{\nu}$). Nonetheless, Bocchino & Bykov (2001) have persisted in suggesting that the systematic errors in the EGRET location contours could contrive to yield the pulsar (or its nebula) as the source of the GeV emission. This is unlikely in our view since an *independent* analysis of >1GeV EGRET photons by Lamb & Macomb (1997) also yields a GeV source, GeV J0617+2237 at the same location as 3EG J0617+2238, both completely inconsistent with the plerion location. So we appear to have come full circle since Sturner & Dermer (1995): unless unrealistically low magnetic fields are adopted, it is *not* possible to reconcile the intensity of the $\gamma$-ray emission with $e^-$ bremsstrahlung or IC models, *given the known radio and X-ray intensities of the region of remnant enclosed by the 95% location contours of 3EG J0617+2238*. In any case, the radio and X-ray emission from the region of the remnant enclosed by the 95% confidence location contours of 3EG J0617+2238 is known to be predominantly thermal. We echo the suggestion of Sturner & Dermer (1995) that IC443 is indeed a hadronic CR accelerator, and the $\gamma$-ray emission is due mostly to pion decay. Using equations 1 & 2, we calculate that just ~30% of the ambient ~7900 $M_o$ (measured via CO) is sufficient to yield a pion decay $\gamma$-ray flux consistent with the measured value: $F\gamma(>100 MeV) = (51.5 \pm 3.4) \times 10^{-8}$ photons cm$^{-2}$ sec$^{-1}$ (Hartman et al., 1999).

## CONCLUSIONS

In the future, orbiting GeV telescopes such as GLAST will have the spatial resolution, and the efficiency to both precisely trace these molecular beam dumps, and to (presumably) verify, in a statistically significant manner, the signature hump of pion produced $\gamma$-rays at ~67.5 MeV at CR acceleration sites. This feature has thusfar only been conclusively seen in the diffuse galactic emission (Hunter et al., 1997) and possibly also in the spectra of the EGRET sources towards the galactic center (Markoff et al., 1999), & towards some SNRs. When they come on-line, km scale neutrino telescopes will also be important in the hunt for the hadronic CR sources, together with the next-generation stereo TeV Čerenkov telescopes, which will be able to accurately probe the maximum energy of the accelerated particles. The importance of molecular clouds cannot be overemphasized in the search for hadronic CR sources: one may have a CR "beam", but a dense target medium for the freshly accelerated CRs will tremendously amplify the pion-produced fluxes of high energy photons and neutrinos. (In the case of neutrinos, however, photomeson processes eg., $p + \gamma \rightarrow X + \nu$, will also contribute copious neutrinos if dense photon fields exist at the CR source sites).

Y.M.B. thanks the organisers of the Moriond conference for the excellent alpine venue. He requests, however, that they endeavor to improve the snow conditions for future conferences. We thank Ron Murphy and Chuck Dermer for letting us use their Figure 6 from Murphy, Dermer & Ramaty, 1987. We are also grateful to Tom Gaisser, Ray Protheroe & Todor Stanev for letting us reproduce Fig. 3 from their 1998 paper. The *EGRET* and HEASARC archives at the Goddard Space Flight Center; the on-line Australia Telescope National Facility (ATNF) pulsar archive; as well as the *ROSAT* all-sky survey from the Max-Planck Institute were all invaluable to this study. The Relativistic Astrophysics Group at *IAR* (JAC & GER) is supported by *CONICET*, *ANPCT*, and Fundación Antorchas. YMB is supported by the High Energy Astrophysics Division and NASA contract NAS8-39073.